\documentclass[prA,floatfix,amsmath,twocolumn]{revtex4}

\usepackage{amsthm}
\usepackage{bbm}
\usepackage{amssymb}
\usepackage{amsmath}

\newtheorem{thm}{Theorem}
\newtheorem*{thm*}{Theorem}

\newtheorem{lem}[thm]{Lemma}
\theoremstyle{definition}

\newtheorem{obs}[thm]{Remark}

\theoremstyle{remark}

\newcommand{\h}{\mathcal{H}}
\newcommand{\1}{\mathbbm{1}}

\renewcommand{\>}{\rangle}
\newcommand{\<}{\langle}

\DeclareMathOperator{\trace}{tr}

\DeclareMathOperator{\vol}{vol}
\DeclareMathOperator{\dist}{d}

\begin{document}

\title{Optimality of programmable quantum measurements}

\author{D. P{\'e}rez-Garc{\'\i}a}
\affiliation{Max Planck Institut f{\"u}r Quantenoptik,
Hans-Kopfermann-Str. 1, Garching, D-85748, Germany}
\affiliation{{\'A}rea de Matem{\'a}tica Aplicada, Universidad Rey
Juan Carlos, C/ Tulipan s/n, $28933$ M{\'o}stoles (Madrid), Spain}

\begin{abstract}
We prove that for a programmable measurement device that
approximates {\it every} POVM with an error $\le \delta$, the
dimension of the program space has to grow at least polynomially
with $\frac{1}{\delta}$. In the case of qubits we can improve the
general result by showing a linear growth. This proves the
optimality of the programmable measurement devices recently
designed in [G. M. D'Ariano and P. Perinotti, Phys. Rev. Lett.
\textbf{94}, 090401 (2005)].

\end{abstract}

\date{February 8, 2006}

\maketitle

\section{Introduction}

One of the most important features of nowadays computers is their universality. That
is, the same computer can achieve many different tasks by changing the program that
runs in it. The analogue quantum concept -the search for universal quantum devices- has
attracted a lot of attention along the short history of quantum information and quantum
computation. The idea behind is the same as in the classical situation: the economy of
resources. Given the difficulty of creating a quantum device, it would be desirable to
create it as universal (or multipurpose) as possible. Most of the work in this
direction has been concentrated in the two basic quantum operations: unitaries (or more
generally channels) \cite{NCh}-\cite{HZB3} and measurements \cite{DB}-\cite{MaPe1}.

\

In this paper we will restrict ourselves to the latter.  That is,
we will study measurement apparatus that can be programmed to
achieve any generalized measurement (POVM). The program will be an
ancillary quantum state that can be changed depending on the POVM
one wants to get. The possible applications of these kind of
devices are considerably large: measurement based quantum
computation, eavesdropping of quantum encrypted information, and
in general every quantum protocol in which one wants to change the
measurements on the fly. However, exact universally programming of
measurements is impossible \cite{DB}, \cite{FDF} as a consequence
of the no-go theorem for programmability of unitary
transformations \cite{NCh}. Hence, one has to restrict to schemes
that approximate any measurement with a fixed error $\delta$.

\

The first universal programmable quantum device was designed in
\cite{FDF}, but it needed a dimension $m$ of the ancilla that
grows exponentially versus the inverse of the error $\delta^{-1}$.
This  was dramatically improved in \cite{MaPe1}, where only a
polynomial (and linear in the case of qubits!) growth of $m$ is
required. In this paper we will show that the results in
\cite{MaPe1} are optimal, by showing that:
\begin{enumerate}
\item The polynomial growth cannot be improved, that is, the dimension of the ancillary
system has to scale at least like
$m\succ\left(\frac{1}{\delta}\right)^{\frac{d-1}{2}}$, where $d$ is the dimension of
the original system. \item The linear growth cannot be improved in the case of qubits,
that is, we have at least the scaling $m\succ\frac{1}{\delta}$.
\end{enumerate}

It is quite surprising that the estimate for the general case does
not give the optimal exponent in the case of qubits. The reason
comes from the techniques used in the paper. Since they come from
the theory of convex bodies, we need a \textit{real} vector space.
This is straightforward in the case of qubits, where one has the
Bloch sphere representation, but a much more artificial trick has
to be used in the general case.

\section{Preliminaries}

Let us recall that in quantum mechanics, the statistics of a generic measurement
apparatus (with discrete sampling space) is described by a positive operator valued
measure (POVM), that is, a set of positive operators (one for each possible outcome)
$P^j\ge 0$ on the Hilbert space of the system such that $\sum_j P^j=\1$. The statistics
of the outcomes $j$ for an input state $\rho$ are given by the Born rule
    $$p(j|\rho)=\trace(\rho P^j).$$

    How to design then a programmable POVM? The idea is to build a device that acts on
    the original system (we will call its dimension $d$) and on some ancillary or program
    system (with dimension $m$) that can be tuned just by changing the state ("quantum
    software") in the ancilla. Clearly, the most general programmable measurement
    device would be a fixed unitary $U$ in both the system and the ancilla, followed by
    a POVM $(F^j)_j$ in the joint system. That is, if the ancilla state is $\sigma$, we
    will get the statistics
        $$p(j|\rho)=\trace(U(\rho\otimes \sigma)U^{\dagger}F^j)\quad \forall j,\rho.$$
        Including the unitary $U$ in the POVM $(F^j)^j$ restricts our study to programmable
        POVMs
        $(G^j)_{j=1}^d$
        of the form
            \begin{equation}
    \label{eq.G}G^j=\trace_a((\1\otimes \sigma)F^j),
            \end{equation}
     where $(F^j)_{j=1}^d$ is a
        POVM in the joint system.

        \

        And how to {\it measure} the distance between the
        original POVM $\mathbb{P}=(P^j)_{j=1}^d$ and the programmed one
        $\mathbb{G}=(G^j)_{j=1}^d$? Maybe the most natural measure is given by the usual
        distance between the probability distributions of the outcomes; in our case
        $$\dist(\mathbb{P},\mathbb{G})=\max_{\rho}\sum_{j=1}^d\left|\trace(\rho(P^j-G^j))\right|.$$
            Clearly $\max_j\|P^j-G^j\|_\infty\le \dist(\mathbb{P},\mathbb{G})\le
            d\max_j\|P^j-G^j\|_\infty$, and since $\frac{1}{d}\|\cdot\|_1\le
            \|\cdot\|_\infty \le \|\cdot
            \|_1,$ and $d$ is a fixed constant (we are just interested in the asymptotics
            in $m$), we will be able to reason also with $\|\cdot\|_1$.

            \

            So we want to construct a POVM $(F^j)_{j=1}^d$ in the joint system such
            that, for any POVM $(P^j)_{j=1}^d$ in the original system, there exists an
            ancillary state $\sigma$ such that $\dist(\mathbb{G},\mathbb{P})\le \delta$,
            where $\mathbb{G}=(G^j)_j$ is given by (\ref{eq.G}). It is argued in
            \cite{MaPe1} that it is enough to approximate POVMs that are given by
            $1$-dimensional orthogonal projectors $P^j=|\phi^j\>\<\phi^j|$. We will
            call such a POVM $(F^j)_{j=1}^d$ a $\delta$-universal programmable
            measurement ($\delta$-UPM).

\section{Results}

            The first result of this paper is to relate
            the dimension of the ancillary system with the existence of disjoint nets
            of balls in some Hilbert space (we will prove it at the end of the paper).

            \begin{thm}\label{thm1}
            For a $\delta$-UPM, the dimension $m$ of the ancillary system verifies
 $m\ge\frac{1}{d} A$, where $A$ is the cardinality of a net of pure states
$(|\phi_\alpha\>)_{\alpha=1}^A$ in a $d$-dimensional Hilbert space such that
\begin{equation}\label{eq.4}
D(|\phi_\alpha\>,|\phi_\beta\>)>\sqrt{8d\delta}
\end{equation}
for any $\alpha\not =\beta$, where \begin{equation}\label{eq.0}
D(|\phi\>,|\varphi\>)=\||\phi\>\<\phi|
-|\varphi\>\<\varphi|\|_1=\sqrt{1-|\<\phi|\varphi\>|^2}.
\end{equation}
 \end{thm}
\

Then, to establish lower bounds for the dimension of the ancilla $m$, it is enough to
obtain lower bounds for $A$.

\subsection*{The case of qubits}

Let us start with the case of qubits. The key estimate will be given by the following
lemma.

\begin{lem}\label{lem2}
Let us consider a small $\epsilon$ $(0<\epsilon<\frac{1}{10})$. Then, in the unit
sphere $S$ of a $n$-dimensional real Hilbert space $\h$, one can take
$\frac{1}{(10\epsilon)^{n-1}}$ elements $x_j$ with the property that
$\|x_j-x_i\|>\epsilon$ if $j\not = i$.
\end{lem}

\begin{proof}
Let us consider a maximal subset $(x_j)_{j=1}^J$ such that $\|x_j-x_i\|\ge
2\epsilon>\epsilon$  if $j\not = i$. By maximality, $S$ can be covered by balls of
radius $2\epsilon$ centered in the $x_j$'s. Moreover, if $\|x\|\ge 1-\epsilon$, we have
that there exists a $j$ such that $\|\frac{x}{\|x\|}-x_j\|\le 2\epsilon$ and so
$$\|x-x_j\|\le 2\epsilon +\|x-\frac{x}{\|x\|}\|\le 3\epsilon.$$
This means that, if $B$ is the unit ball of $\h$, the set
$C=\bigcup_{j=1}^J\{x_j+3\epsilon B\}$ covers the ring $R=\{x\in \h: 1-\epsilon\le
\|x\|\le 1\}$. Therefore,
$$J3^n\epsilon^n\vol(B) \ge \vol(C)\ge \vol(R)=\left(1-(1-\epsilon)^n\right)\vol(B),$$
and so $$J\ge
\frac{1}{3^n}\left(\frac{1}{\epsilon^n}-\left(\frac{1}{\epsilon}-1\right)^n\right)\ge$$
$$\ge \frac{1}{3^n}\left(\frac{1}{\epsilon}-1\right)^{n-1}\ge
\frac{1}{(10\epsilon)^{n-1}} .$$
\end{proof}

Now we consider the Bloch sphere. The distance $D$ of (\ref{eq.0}) corresponds to
$\frac{1}{2}$ the usual (Hilbert) distance in the Bloch sphere. Therefore, using Lemma
\ref{lem2} (now $n=3$), we can take a net of $\frac{1}{6400}\frac{1}{\delta}$ pure
states with property (\ref{eq.4}), which immediately implies that the dimension $m$
that we need in the ancilla to get a $\delta$-UPM has to verify $$m\ge
\frac{1}{12800}\frac{1}{\delta}.$$

That is, the linear growth obtained in \cite{MaPe1} is optimal.

\begin{obs}\label{rem1}
Since the main aim under study is the growth of $m$ with $\frac{1}{\delta}$, we are
quite careless with the constants and then, as one can easily see, the constant
$\frac{1}{12800}$ is far from optimal.
\end{obs}

\subsection*{The general case}

Let us now turn to the general case. Since we do not have now such a good real
representation we will play the trick of restricting to some {\it real part} $\h_R$ of
our $d$-dimensional Hilbert space $\h$, that is, we fix one orthonormal basis $|i\>$
and we consider the elements that are of the form $\sum_{i=1}^d \lambda_i |i\>$ with
the $\lambda_i$'s real. This is a $d$-dimensional real Hilbert space with the inherited
norm (given by $\left(\sum_i|\lambda_i|^2 \right)^\frac{1}{2}$).

\

Now we also have to {\it identify} vectors up to global phases. For this reason we
consider in the unit sphere of $\h_R$ a maximal set $X=(x_j)_{j=1}^J$ with the
following two properties:
\begin{enumerate}
\item[(P1)] $\|x_j-x_i\|\ge 2\epsilon$ if $j\not =i$, \item[(P2)]
if $x\in X$, then $-x\in X$.
\end{enumerate}

We have that the unit sphere $S_R$ of $\h_R$ can be covered by balls of radius
$2\epsilon$ centered in the $x_j$'s. Let us see it:

If it is not the case, there exists an $x\in S_R$ such that $\|x-x_j\|>2\epsilon$ for
every $j$. By (P2), also $\|-x-x_j\|>2\epsilon $ for every $j$, and this implies that
$X\cup \{x,-x\}$ also fulfils (P1) and (P2), which contradicts the maximality of $X$.
Then, reasoning as in Lemma \ref{lem2} we get that $J\ge \frac{1}{(10\epsilon)^{d-1}}$.

Now we choose either $x$ or $-x$ for every $x\in X$ and obtain another sequence
$(|\phi_j\>)_{j=1}^{\frac{J}{2}}$, that one can see in the original Hilbert space $\h$,
for which, if $j\not =i$, both $$\||\phi_j\>-|\phi_i\>\| \quad \text{ and } \quad
\||\phi_j\>+|\phi_i\>\| \ge 2\epsilon.$$ Therefore,
$$D(|\phi_j\>,|\phi_i\>)= \sqrt{1-|\<\phi_j|\phi_i\>|^2}\ge
{1-|\<\phi_j|\phi_i\>|}= $$
$$= \min\left\{{1-\<\phi_j|\phi_i\>},{1+\<\phi_j|\phi_i\>}\right\}=$$
$$=\frac{1}{2}
\min\left\{\left\||\phi_j\>-|\phi_i\>\right\|,\left\||\phi_j\>+|\phi_i\>\right\|\right\}\ge
\epsilon.
$$

In conclusion, we can obtain property (\ref{eq.4}) with $A\ge
k(d)\left(\frac{1}{\delta}\right)^\frac{d-1}{2}$ (now
$\frac{1}{k(d)}=2(20\sqrt{8})^{d-1}d^\frac{d-1}{2}$ is again far from optimal), which
implies that the dimension of the ancilla $m$ needed to get a $\delta$-UPM has to be
$$m\ge k'(d) \left(\frac{1}{\delta}\right)^\frac{d-1}{2},$$ ($k'(d)=\frac{1}{d}k(d)$).

So the best growth for $m$ is  polynomial in $\frac{1}{\delta}$. This implies that the
control unitary (which has polynomial growth \cite{MaPe1}) is essentially optimal among
the programmable quantum measurements.

\subsection*{The proof of the Theorem}

Let us finish the paper with the proof of Theorem \ref{thm1}. We will need the
following lemma.

\begin{lem}\label{lem1}
If $0\le \lambda_i\le 1$ for every $1\le i\le I$ and
\begin{equation}\label{eq.1}
\left\|\sum_{i=1}^I\lambda_i |\psi_i\>\<\psi_i|-|\phi\>\<\phi|\right\|_1\le \epsilon,
\end{equation}
then there exists an $i_0$ such that $D(|\phi\>,|\psi_{i_0}\>)\le
\sqrt{2\epsilon}$.
\end{lem}

\begin{proof}
Taking trace in (\ref{eq.1}) we have that
$\lambda=\sum_i\lambda_i$ verifies $|\lambda-1|\le \epsilon$. By
defining $\tilde{\lambda}_i=\frac{\lambda_i}{\lambda}$ we get that
$\sum_i\tilde{\lambda}_i=1$ and still
\begin{equation*}
\left\|\sum_i\tilde{\lambda}_i
|\psi_i\>\<\psi_i|-|\phi\>\<\phi|\right\|_1\le 2\epsilon.
\end{equation*}

Now
$$1-\sum_i\tilde{\lambda}_i|\<\phi|\psi_i\>|^2=\left|\<\phi|\left(\sum_i\tilde{\lambda}_i
|\psi_i\>\<\psi_i|-|\phi\>\<\phi|\right)|\phi\>\right|$$ $$\le
\left\|\sum_i\tilde{\lambda}_i
|\psi_i\>\<\psi_i|-|\phi\>\<\phi|\right\|_\infty \le 2\epsilon$$
which means that there exists an $i_0$ with
$|\<\phi|\psi_{i_0}\>|^2\ge 1-2\epsilon$; and hence
$D(|\phi\>,|\psi_{i_0}\>)\le \sqrt{2\epsilon}$.
\end{proof}

Now we want to choose a programmable POVM $(F^j)_{j=1}^d$ that approximates any
observable $(|\phi^j\>\<\phi^j|)_{j=1}^d$ with an error $\le \delta$. This means that
$$
\dist((G^j)_j,(|\phi^j\>\<\phi^j|)_j)=\max_\rho\sum_j\left|\trace[\rho(G^j-|\phi^j\>\<\phi^j|)]
\right|\le \delta,
$$
where $G^j=\trace_a((\1\otimes |\varphi\>\<\varphi|)F^j)$ for some
$|\varphi\>$ (by convexity it is enough to consider pure states in
the ancilla).

\

This will immediately give us that, in particular, $F=F^1$ has to
approximate $|\phi\>\<\phi|$ with error $\le \delta$ in the
$\infty$-norm for any arbitrary pure state $|\phi\>$ . Using that
$\frac{1}{d}\|\cdot\|_1\le \|\cdot\|_\infty\le \|\cdot\|_1$ we
will obtain precision $d\delta$ in the $1$-norm. That is, if we
take a sequence $(|\phi_\alpha\>)_{\alpha=1}^A$ of pure states
with property (\ref{eq.4}) (as in the statement of Theorem
\ref{thm1}), we have that for every $\alpha$ there exists a
$|\varphi_\alpha\>$ such that $\|\rho_\alpha-
|\phi_\alpha\>\<\phi_\alpha|\|_1\le d\delta$, where $\rho_\alpha=
\trace_a((\1\otimes |\varphi_\alpha\>\<\varphi_\alpha|)F)$.

Now we take the spectral decomposition
$F=\sum_{i=1}^I\lambda_i|\tilde{\psi}_i\>\<\tilde{\psi}_i|$ ($0\le
\lambda_i\le 1$). Theorem \ref{thm1} will be then proven if we can
show that
$$A\le I.$$

To see it let us fix an $\alpha$. Calling
$|\psi_i^\alpha\>=\<\varphi_\alpha|\tilde{\psi}_i\>$ we have that
$\rho_\alpha=\sum_i \lambda_i|\psi_i^\alpha\>\<\psi_i^\alpha|$. By
the hypothesis in $F$ and Lemma \ref{lem1} we have that there
exists an $i_\alpha$ such that
\begin{equation}\label{eq.3}
D(|\phi_\alpha\>,|\psi_{i_\alpha}^\alpha\>)\le \sqrt{2d\delta}.
\end{equation}

Now, for $\alpha\not =\beta$,
$$D(|\tilde{\psi}_{i_\alpha}\>,|\tilde{\psi}_{i_{\beta}}\>)\ge D(|\phi_\alpha\>|\varphi_\alpha\>,
|\phi_{\beta}\>|\varphi_{\beta}\>)-$$ $$-
D(|\tilde{\psi}_{i_\alpha}\>,|\phi_\alpha\>|\varphi_\alpha\>)-
D(|\tilde{\psi}_{i_{\beta}}\>,|\phi_{\beta}\>|\varphi_{\beta}\>)>0.$$

To see it, it is enough to notice that, by (\ref{eq.3}), both
$D(|\tilde{\psi}_{i_\alpha}\>,|\phi_\alpha\>|\varphi_\alpha\>)$ and
$D(|\tilde{\psi}_{i_{\beta}}\>,|\phi_{\beta}\>|\varphi_{\beta}\>)$ are bounded by
$\sqrt{2d\delta}$; and by (\ref{eq.4}), $D(|\phi_\alpha\>|\varphi_\alpha\>,
|\phi_{\beta}\>|\varphi_{\beta}\>)>\sqrt{8d\delta}$.

Therefore, if $\alpha\not =\beta$, we have that $|\tilde{\psi}_{i_\alpha}\>\not
=|\tilde{\psi}_{i_{\beta}}\>$, which means that $i_\alpha\not =i_\beta$; and hence
$A\le I$. QED.

\section{Conclusion}

In conclusion, we have proven that the universal programmable measurements designed in
\cite{MaPe1} are optimal in the sense of the resources (dimension of the program space)
needed to built them. This opens the door of a key question: how to physically
implement them?

\

Another question that arises from this paper, apart from improving the constants (see
Remark \ref{rem1}), is to fill in the gap between the lower and the upper bounds
\cite{MaPe1} found for the exponents in the general case: $\frac{d-1}{2}\le \text{
exponent }\le d(d-1)$. Notice that in the case of qubits the optimal exponent $1$ does
not coincide with any of the general bounds.

\

As for the techniques, this paper shows once more the close
connection between Quantum Information and the rich mathematical
theory of convex bodies (other recent applications can be found
for instance in \cite{HLW}-\cite{SBZ} and the references therein).
Our believe is that this connection will give much more in the
near future.

\section*{Acknowledgments}

The author would like to thank M. M. Wolf, J. I. Cirac and
specially P. Perinotti and G. M. D'Ariano for valuable discussions
concerning this paper. This work has been  partially supported by
Spanish grant MTM-2005-0082.

\end{document}